\documentstyle[12pt]{article}

\title{\bf Geometric phase in phasing of antenna arrays }

\vspace{60mm}

\author{\bf Rajendra~Bhandari}

\date{ }

\begin{document}

\maketitle
\vspace{10mm}
\begin{center}
\begin{tabular}{ll}
            & Raman Research Institute, \\
            & Bangalore 560 080, India. \\
            & email: bhandari@rri.ernet.in\\           
\end{tabular}
\end{center}
\vspace{40mm}
\begin{abstract}
The response of a pair of differently polarized antennas is determined
by their polarization states {\it and} a phase between them which has a 
geometric part which becomes discontinuous at singular points in
the parameter space. Some consequences are described.\\
\end{abstract}

-----------------------------------------------------------------------\\
Submitted for Proceedings, IAU199 Symposium : The Universe at low radio 
frequencies, Pune, India, Nov.30-Dec.4, 1999.

\newpage

{\bf Introduction :} The geometric phase 
(also popularly known as Berry's phase) finds
some of its most easily visualizable manifestations in the
physics of polarized light \cite{rbreview}. In 1956, Pancharatnam defined the
`in-phase' condition for two different, non-orthognal 
polarization states to be one for which their interference yields maximum
intensity 
and discovered that under a cycle of transformations
of the polarization state along a closed geodesic polygon on the
Poincar\'{e} sphere (PS) the beam acquires a phase equal to half the solid
angle subtended by the polygon at the centre. Further work by the 
present author has shown that the above geometric phase 
exhibits measurable jumps at singular points in 
the parameter space such that a circuit around the singularity 
results in a measurable phase shift equal to $2n\pi$ \cite{rbdirac,rbreview}.
The flat behaviour of the 
phase near a singularity has been used in adaptive optics 
to make a spatial light
modulator for pure intensity modulation, keeping phase constant \cite{love}. 
In arrays 
phased by geometric phase shifters \cite{fox}, phase singularities 
lead to the possibility of an array looking in two different 
directions at two different wavelengths \cite{qhq}.\\

\begin{figure}
\vspace{3.3in}
\caption{Two elliptically polarized antennas in states A and P, 
in phase when $\phi=0$, get out of phase when $\phi \neq 0$, by 
an amount shown in the curves on the right.}
\end{figure}

\newpage

{\bf A pair of antennas with different polarization :}
Take two identical elliptically polarized
antennas, in phase with each other, so that their resultant intensity 
response is maximum `on-axis'. Now rotate one of the antennas with
respect to the other by an angle $\phi/2$ (figure 1). The two will  no 
longer be in phase in that their combined response will not be
maximum `on-axis'. The phase difference $\psi$ between them
(Pancharatnam), given by $tan\psi=cos\theta~ tan(\phi/2)$,
is shown in figure 1 for a few values of the polar angle $\theta$ 
of the states on the PS. Note 
the phase shift  (i) is of magnitude $\pi$ for a $2\pi$ rotation 
on the PS and (ii) jumps through 
$\pm \pi$ near $\theta=90^\circ,~ \phi=180^\circ$ (as happens for 
$2\pi$ rotation in real space of 
particles with odd half integer spin, verified in analogous 
polarization experiments in an optical interferometer \cite{4pism}. 
We note that a similar behaviour is implicit in the 
special case ${\theta}_1={\theta}_2$,
Q=V=U=0 of equation (8) in Morris, Radhakrishnan Seielstad \cite{morris}
\footnote{This equation has been re-derived by Weiler \cite{weiler} and  
Nityananda \cite{rn}.}.\\

{\bf Interference nulls for non orthogonal states:}
When radiation from a partially polarized source with degree of
polarization p and eigenpolarizations $P$ and $\tilde P$ is 
picked up by two antennas tuned to polarizations  $A_1$ and $A_2$,
then for every polarization state $A_1$, there is a state $A_2$, 
{\it not orthogonal to} $A_1$, such that the correlation
of the two outputs is zero \cite{radpp}. A simple way
to prove this curious result is to consider a superposition of two
interference patterns; (i) due to a fraction $(1+p)/2$ of the radiation
in state $P$ and (ii) due to a fraction $(1-p)/2$ in state $\tilde P$,
with a phase difference of magnitude $\pi$, which is 
a geometric phase due to  the surface enclosed by the closed 
geodesic ~~curve $P$ $A_1$ $\tilde P$ $A_2$ $P$ on the PS (a hemisphere).


\newpage

\begin{thebibliography}{4pism}
\addcontentsline{toc}{section}{References}
\bibitem{rbreview}
Bhandari R. 1997, Phys. Rep., 281, 1
\bibitem{rbdirac}
Bhandari R. 1991, Phys. Lett. A, 157, 221
\bibitem{love}
Love G. D., Gourlay J. 1996, Opt. Lett., 21, 1496
\bibitem{fox}
Fox A. G. 1947, Proc. I.R.E., 35, 1489
\bibitem{qhq}
Bhandari R. 1995, Phys. Lett. A, 204, 188
\bibitem{4pism}
Bhandari R. 1993, Phys. Lett. A, 180, 15
\bibitem{morris}
Morris D., Radhakrishnan V.,  Seielstad G. A. 1964,
Ap.J., 139, 551
\bibitem{weiler}
Weiler K. W. 1973, Astron.  Astrophys., 26, 403
\bibitem{rn}
Nityananda R. 1994, Current Science, 67, 243
\bibitem{radpp}
Radhakrishnan V. 1994, Current Science, 67, 257


\end{thebibliography}
\end{document}